\documentclass[manuscript,screen]{acmart}
\AtBeginDocument{%
  \providecommand\BibTeX{{%
    \normalfont B\kern-0.5em{\scshape i\kern-0.25em b}\kern-0.8em\TeX}}}

\setcopyright{acmcopyright}
\copyrightyear{2018}
\acmYear{2018}
\acmDOI{XXXXXXX.XXXXXXX}

\acmConference[SIGCSE TS]{Technical Symposium on Computer Science Education}{20–23 March,
  2024}{Portland, Oregon, United States}
\acmPrice{15.00}
\acmISBN{978-1-4503-XXXX-X/18/06}

\begin{document}
\title[Redefining Computer Science Education with AI-Based No-Code Platforms]{
Redefining Computer Science Education: Code-Centric to Natural Language Programming with AI-Based No-Code Platforms
%Programming with AI and it's implications on Computer Science Education
}


%\author{Anonymous}
\author{David Y.J. Kim}
\email{dyjkim@mit.edu}
\orcid{0000-0003-4057-0027}
\affiliation{%
  \institution{Massachusetts Institute of Technology}
  \streetaddress{77 Massachusetts Ave}
  \city{Cambridge}
  \state{MA}
  \country{USA}
  \postcode{02139}
}

\renewcommand{\shortauthors}{David Y.J. Kim}

\begin{abstract}
This paper delves into the evolving relationship between humans and computers in the realm of programming. 
Historically, programming has been a dialogue where humans meticulously crafted communication to suit machine understanding, shaping the trajectory of computer science education. 
However, the advent of AI-based no-code platforms is revolutionizing this dynamic. 
Now, humans can converse in their natural language, expecting machines to interpret and act. 
This shift has profound implications for computer science education. 
As educators, it's imperative to integrate this new dynamic into curricula. 
In this paper, we've explored several pertinent research questions in this transformation, which demand continued inquiry and adaptation in our educational strategies.
  % A clear and well-documented \LaTeX\ document is presented as an
  % article formatted for publication by ACM in a conference proceedings
  % or journal publication. Based on the ``acmart'' document class, this
  % article presents and explains many of the common variations, as well
  % as many of the formatting elements an author may use in the
  % preparation of the documentation of their work.
\end{abstract}





\received{20 February 2007}
\received[revised]{12 March 2009}
\received[accepted]{5 June 2009}

\maketitle

\section{Introduction}
Automatic programming~\cite{mur2006automatic,Bodk2008ProgramSB,Basin2004SynthesisOP}, with its rich historical tapestry, has always been about bridging the communication gap between human intent and machine execution. 
At its core, it embodies the principle of transforming higher-level instructions into forms that a computer can readily understand and act upon.
Today's automatic programming aims to streamline the software development process, enabling developers to emphasize problem definition over intricate implementation details. 
This evolution promises a future of enhanced collaboration between humans and machines in software creation, making the process more accessible, efficient, and resilient against errors.

The advent of artificial intelligence (AI) and machine learning (ML) technologies has catalyzed a significant evolution in automatic programming~\cite{lunn2021computer,Rietz2021CodyAA,puri2021codenet}. 
Modern automatic programming systems leverage AI and ML to generate code from a higher-level specification or even from natural language descriptions~\cite{Kumar2023OpenAC}. 
This has broadened the scope of automatic programming from simple code translation to complex code synthesis, opening up new possibilities for software development~\cite{Yellin2023ThePO,Bull2023GenerativeAA}.

Take, for instance, GitHub Copilot, a coding assistant powered by OpenAI Codex~\cite{chen2021evaluating}. 
GitHub Copilot offers contextually relevant code suggestions as you type, effectively automating part of the coding process. 
Whether you're learning a new language or building a complex system, Copilot serves as a knowledgeable companion, aiding in writing new code, navigating existing code, and even generating unit tests~\cite{Nguyen2022AnEE,Barke2022GroundedCH,Denny2022ConversingWC,Dakhel2022GitHubCA,Imai2022IsGC,Yetistiren2022AssessingTQ}.
Another example is CodeWhisperer, an AI-powered code generation tool~\cite{Yetistiren2023EvaluatingTC}. 
Amazon CodeWhisperer is a general purpose, machine learning-powered code generator that provides you with code recommendations in real time. 
As you write code, CodeWhisperer automatically generates suggestions based on your existing code and comments.

MIT Aptly is also a cutting-edge tool leveraging large language model technology to automatically generate mobile applications from natural language descriptions~\cite{Kim2022SPEAKYM,Granquist2023AIAF}. 
Built on OpenAI's Codex, the same foundation as GitHub's Copilot, Aptly enables the creation of functional software without \textbf{any need for traditional coding}. 
Users provide the tool with a description of the desired app, and Aptly generates the corresponding mobile application. 
This technology has significant implications for both professional programming and computer science education.
AI-based no-code platforms such as Aptly challenges the traditional emphasis on text-based or block-based coding in the current curriculum, prompting a reevaluation of how programming is taught when the transition from ideas to functioning programs can be automated.

Automatic programming promises to significantly transform the future of software engineering, some go far as to say that it is the end of programming~\cite{Welsh2022TheEO}. 
As it advances, developers may spend less time on routine coding tasks and more time on high-level problem-solving, design, and innovation. 
The reduced need for manual coding could accelerate software development processes and minimize human errors, leading to more reliable and efficient software. 
The democratization of programming, facilitated by the lower technical barriers to entry, could also foster greater diversity in the field, resulting in a broader range of perspectives and solutions.

\begin{figure*}[h]
  \centering
  \includegraphics[width=\linewidth]{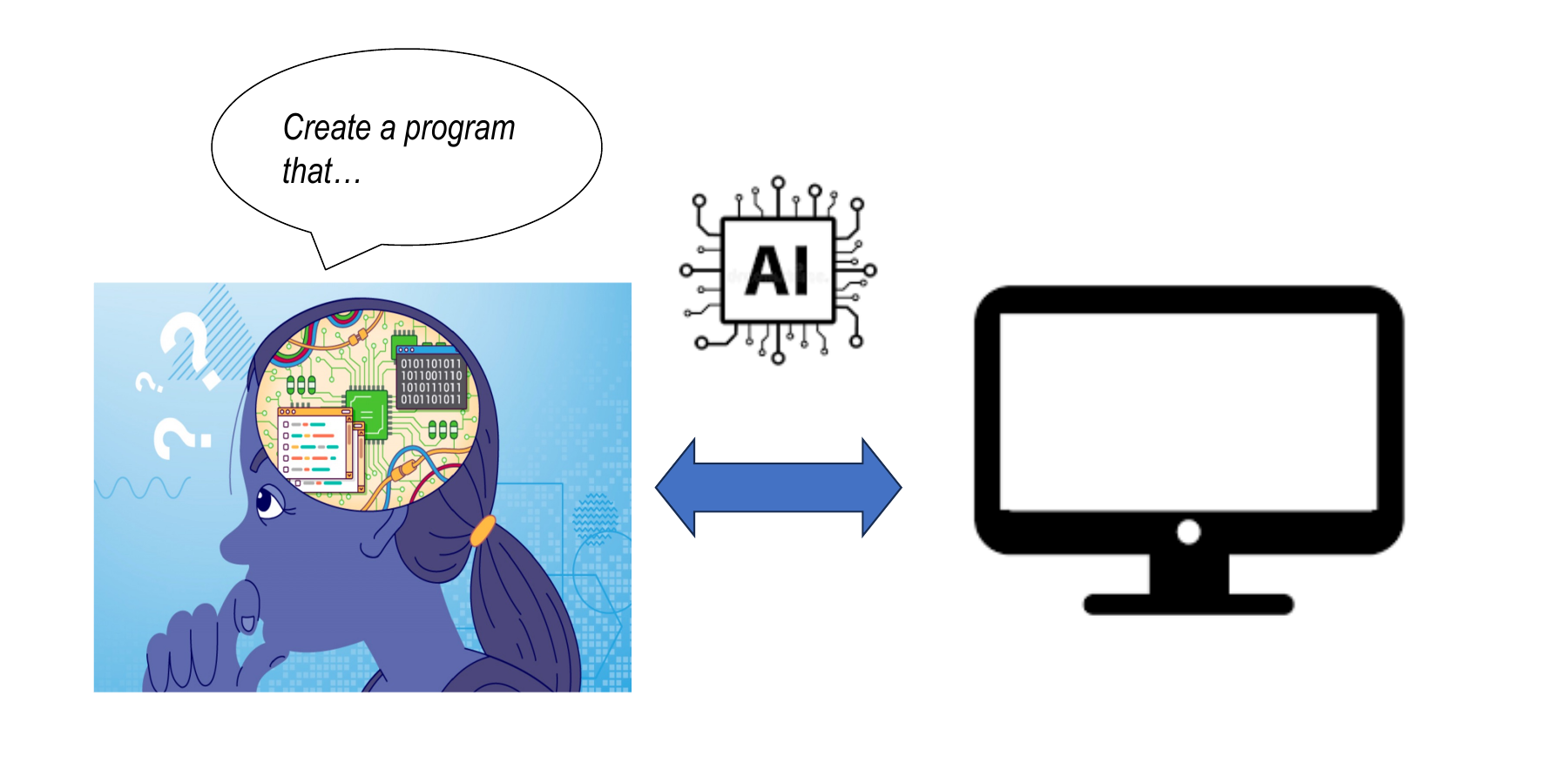}
  \caption{In its essence, programming can be seen as a form of dialogue or conversation between a human and a computer. Traditionally, the onus is largely on the human to code in a way that the computer can understand.  However, with the rise of AI, this dynamic is starting to shift as the conversation is becoming more human-centric}
\end{figure*}

% Automatic programming brings numerous benefits to the software development process. One of the most significant is the reduction in the amount of manual coding needed. By automating the generation of routine or repetitive code, automatic programming allows developers to focus on more complex and creative aspects of software development, increasing productivity and efficiency.
% For example, automatic programming can help minimize human error in code. Since a significant portion of code is generated automatically, the likelihood of mistakes stemming from manual coding—such as syntax errors, incorrect variable usage, or wrong function calls—is greatly reduced. This can lead to more reliable and robust software.
% Automatic programming can also enhance the efficiency of software development. By automating time-consuming tasks like writing boilerplate code or setting up basic software infrastructure, developers can focus on designing the unique features and functionalities of their software, leading to faster development cycles.

Despite the significant advancements and benefits that automatic programming brings to the software development process, its influence on computer science education is not as thoroughly examined~\cite{becker2023programming}. 
Traditional computer science education has long been centered on manual coding and algorithm design~\cite{lunn2021computer}. 
It emphasizes the need for students to understand the intricacies of programming languages, the efficiency of algorithms, and the design of complex software systems~\cite{Faherty2023BriefHO,Falkner2019AnIC}. 
AI based code generation techniques, while increasingly prevalent in the industry, often receive less attention in the educational context.

However, as this trend continues to revolutionize the software industry, it's becoming increasingly clear that our approach to computer science education may need to evolve to keep pace with these changes. 
The advent of automatic programming poses both opportunities and challenges for computer science education, which merit careful exploration and discussion.

In this paper, we embark on an in-depth exploration of the potential ramifications of AI-based no-code programming on the contours of computer science education. 
Our examination is structured as follows: We commence by delineating the foundational principles of programming, tracing the evolution of computer science education in tandem with technological advancements. 
Subsequently, drawing from the author's perspective on the subject, we propose a series of research questions. 
These queries are pivotal for the adaptation of computer science education, ensuring it remains responsive to technological shifts and continues to cultivate the next generation of adept programmers.

\section{Background}
In the subsequent section, we'll delve into the core concepts that underpin the world of programming and chart the evolution of computer science education based on these foundational ideas. 

\subsection{Programming: A Conversation between Human and Computer}

\textit{In its essence, programming can be seen as a form of dialogue or conversation between a human and a computer~\cite{repenning2011making}}. 
The programmer communicates their intentions to the computer in a language it can understand (i.e., a programming language), and the computer responds by executing the commands or providing feedback in the form of output or error messages.

The conversation starts when the programmer writes code to instruct the computer to perform a specific task. 
This is analogous to a human asking a question or giving a command in a conversation. 
The programmer must use the correct syntax and semantics in this ``question'' or ``command'' for the computer to understand and carry out the task correctly.
The computer then ``responds'' by executing the code.
If the code is correct and the task is feasible, the computer will carry out the task and generate the expected output. 
This output can be seen as the computer's ``answer'' to the programmer's ``question''.
If the code contains errors or the task is not feasible, the computer will generate error messages. These can be seen as the computer saying, ``I don't understand your question'' or ``I can't do what you're asking'' prompting the programmer to revise their instructions.
The conversation doesn't end here. Programming is an iterative process. The programmer will often revise their code based on the computer's ``responses'' adjusting their instructions until the computer can successfully carry out the desired task. 
This back-and-forth can continue for some time, particularly for complex programs or tasks.

% Just like in a human conversation, effective communication in programming requires both parties to "understand" each other. 
% The programmer must understand the rules of the programming language (the computer's "language"), and the computer must be able to interpret and execute the programmer's instructions correctly.

\subsection{Charting the Progression: The Co-evolution of Computer Languages and Computer Science Pedagogy}
The development of computer science education can be intimately tied to the evolving relationship between humans and machines, particularly as it pertains to the notion of programming as a form of communication. 
Here we trace the historical trajectory of this relationship, focusing on the medium of this communication: programming languages.

\subsubsection{The Primordial Dialogue: Machine and Assembly Languages}
In the nascent stages of computing, the dialogue between humans and machines was direct, albeit rudimentary~\cite{ORegan2021HistoryOP}. 
Programmers used machine language, a series of binary codes, to explicitly instruct computers. 
This form of communication demanded a comprehensive understanding of the machine's internal operations. Assembly language soon offered a slightly more human-readable abstraction, utilizing mnemonic codes, but still necessitated a thorough grasp of the machine's architecture. 
The educational landscape of this era was fundamentally rooted in comprehending the machine's core, with curricula emphasizing the hardware-software nexus~\cite{Larson2008ASB}.

\subsubsection{High-Level Languages: A Paradigm Shift}
The 1950s and 1960s heralded a transformative phase with the advent of high-level languages like FORTRAN, COBOL, and LISP~\cite{Battarel1978DesignOA}. 
No longer were programmers confined to the stringent lexicon of machine language; they could now convey algorithmic logic in terms more akin to human reasoning. 
This evolution marked a shift in computer science education, with curricula emphasizing algorithmic thinking, abstract reasoning, and problem-solving, transcending the intricacies of machine-specific instructions~\cite{Blaisdell1985HowTT}.

\subsubsection{Structured and Object-Oriented Programming: Refining the Discourse}
The subsequent decades saw the rise of structured programming, introducing a more organized, modular approach to coding, exemplified by languages like Pascal and C~\cite{Stroustrup1986CPL}. 
This era reached its zenith with the advent of object-oriented programming (OOP) in languages such as C++ and Java~\cite{Stefik1989ObjectOrientedPT}. OOP allowed for a more sophisticated dialogue with machines, enabling the modeling of real-world scenarios, entities, and their interrelations. 
In response, educational paradigms shifted focus towards design principles, encapsulation, and the broader philosophy of code organization and reusability~\cite{LunaRamrez2014SupportingSP,Yang2015EnhancingOP}.

\subsubsection{Scripting Languages and Rapid Prototyping: The Era of Iterative Dialogue}
Emerging from the structured foundations of the previous era, scripting languages like Python, Ruby, and JavaScript heralded an age of rapid, iterative communication with machines. 
These languages, tailored for swift development and flexibility, empowered programmers to prototype and refine solutions in real-time. 
Correspondingly, computer science education began to prioritize agility, web-centric solutions, and real-world applicability, reflecting the dynamic nature of the industry~\cite{Celani2012CADSA,Warren2001TeachingPU}.

\subsection{The Future Shift in the Dynamics of Programming}
In all of the traditional programming mentioned above, the onus is largely on the human to ``speak'' in a way that the computer can understand. 
The programmer must learn the syntax and semantics of a programming language and must structure their instructions according to the computer's rules. 
In this sense, the human is conforming to the computer's ``language'' in order to facilitate the conversation. 
The programmer writes the code, the computer executes it, and if there are any errors or misunderstandings, the programmer must adjust their code to better fit what the computer can interpret.

However, with the rise of AI based no-code platforms and automatic programming, this dynamic is starting to shift. 
No-code platforms allow users to create software by manipulating natural language or visual elements and defining their interactions, without needing to write traditional code. 
The underlying platform translates these natural language or visual representations into code that the computer can understand.
In this new paradigm, \textbf{the conversation is becoming more human-centric}. 
Instead of the human needing to learn and adapt to the computer's language, the computer is adapting to a more human-friendly mode of interaction. 
The user can express their intentions in a more intuitive, visual manner, and the platform (i.e., AI model) takes on the task of translating these intentions into a form that the computer can understand.

This shift has implications for computer science education. 
If code can be automatically generated, should we still teach students to write code manually? 
Should we shift the focus from coding to understanding and leveraging these automatic programming tools? 
%Or perhaps the key lies in balancing the two: teaching students how to write code, while also exposing them to automatic programming tools so they can understand when and how to use them effectively.

\subsection{Is Coding a Fundamental Component of Computational Thinking?}
The evolution of technology allows us to shift educational focus away from traditional coding towards articulating the desired end-product, the overarching trajectory of programming has consistently been towards abstraction and simplification. 
From assembly to high-level languages, and now to no-code platforms, the ultimate aim has always been to democratize programming, rendering it more accessible and intuitive for a broader audience.
One could argue that AI-based coding represents yet another stride in the ongoing endeavor to democratize programming for a wider audience.
Nevertheless, some contend that the practice of coding is instrumental in nurturing computational thinking~\cite{wing2006computational,grover2013computational,selby2013computational,shute2017demystifying}, as it cultivates a mindset more aligned with computer logic.

A crucial part of computational thinking involves understanding both the capabilities and limitations of computer systems. 
Coding, as a hands-on practice of communicating with computers, plays an instrumental role in fostering this understanding among learners.
When students engage in coding, they are effectively interacting with the computational machinery underlying their code. 
They learn firsthand that computers, while powerful, operate under a set of rules and constraints. 
For example, students quickly realize that computers can execute instructions at incredible speeds, perform complex calculations, and manage large amounts of data, which are feats nearly impossible for humans to perform manually.

However, students also learn that computers are fundamentally rule-based and lack the ability to understand context or ambiguity without explicit programming. 
They are bound by the logic and instructions provided in the code, and they cannot make judgments or assumptions outside of those parameters. 
For instance, a computer cannot inherently interpret the meaning of natural language, understand visual scenes, or make decisions based on abstract concepts, unless explicitly programmed to do so.
Furthermore, students learn about more technical limitations, such as computational complexity and resource constraints. 
They come to understand that not all problems can be solved efficiently by computers, and that certain tasks may require substantial computational resources.

These insights about the possibilities and limitations of computers are essential to computational thinking. 
They help students to become both realistic and strategic in their problem-solving approach. 
They learn to identify problems that are well-suited for computational solutions and also to recognize when computational methods may not be feasible or efficient.

Moreover, this understanding can help students become more effective users of technology. 
With a clear grasp of what computers can and can't do, students can better evaluate the tools, applications, and technologies they use, and can make more informed decisions about when and how to leverage computational solutions.

In essence, coding serves as a practical, hands-on conduit for students to understand the intricacies of computer systems, fostering a deep comprehension of both their potential and their constraints. 
This understanding is a critical component of computational thinking, equipping learners to navigate the digital world effectively and creatively.

\section{Position \& Research Questions}
Given the rapid evolution of technology, it is evident that clinging solely to traditional coding education may not adequately prepare the next generation of programmers for the future. 
Emerging no-code platforms and automatic programming tools are not merely transient trends; they represent significant shifts in the way software development is approached. 
These tools offer notable benefits, such as enhancing productivity, reducing the barrier to entry for programming, and allowing developers to focus on high-level problem-solving and design. 
Consequently, it is vital that the potential of these tools is recognized and incorporated into computer science education.

However, this raises important research questions: How do we balance the use of these no-code platforms with traditional coding education? How do we ensure that we continue to foster computational thinking skills while leveraging the power of these AI tools to enhance the coding experience? 
\emph{It is crucial to investigate these questions to develop a pedagogical approach that integrates both traditional coding skills and the use of modern automatic programming tools.}

The challenge lies in crafting a curriculum that maintains the rigor of computational problem-solving and algorithmic thinking, while also enabling students to harness the power of automatic programming tools.
This balance could empower students to not only understand the intricacies of computer science but also to apply these concepts effectively and efficiently in the real world.
In the following subsections, we propose a few research questions that needs to be investigated in order to achieve such balance.

\subsection{Timing the Introduction of No-Code Platforms in Computer Science Education}
The question of when to introduce no-code platforms is multifaceted. 
On one hand, introducing these tools early in a student's education could make programming less intimidating, potentially attracting a wider range of students to the field. 
By allowing students to create functional software without needing to write traditional code, these platforms can provide a sense of accomplishment and motivation, which are crucial for sustaining interest in computer science.
On the other hand, there's a risk that introducing no-code platforms too early could lead to a superficial understanding of programming concepts.
If students become reliant on these tools without understanding the underlying principles of programming, they may struggle with more advanced programming tasks later on.

Given these considerations, it is important to investigate the optimal timing for introducing no-code platforms in computer science education. 
Central to this inquiry is the question of foundational knowledge: Should students first establish a grounding in basic programming concepts, such as loops, variables, and data structures, before transitioning to AI-driven platforms? 
There's an underlying tension between the immediacy and convenience offered by these platforms and the foundational skills traditionally deemed essential for computational thinking. 
What teaching strategies can effectively balance the use of no-code platforms with the teaching of traditional coding?

Through investigating these questions, we can hope to develop an evidence-based approach to integrating no-code platforms into computer science education. This will enable us to leverage the benefits of these tools while also ensuring that students develop a solid foundation in programming concepts.

\subsection{Are Traditional Assignments Still Valuable for Fostering Computational Thinking?}
The very essence and relevance of conventional computer science assignments come under scrutiny. 
These advanced tools possess the capability to effortlessly tackle and optimize solutions for rudimentary programming challenges.
Such prowess compels us to reevaluate: In an era where a simple prompt like "implement a quick sort algorithm for a list of numbers" can yield immediate results without manual coding, is there enduring merit in having students manually craft various sorting algorithms, or are these exercises becoming antiquated? 
The crux of the matter lies in discerning whether current assignments truly enhance critical thinking and problem-solving skills or if they merely engage students in replicable coding tasks that are rapidly becoming automatable.

With the advent of calculators, certain mathematical abilities, notably swift manual calculations, were rendered largely obsolete in educational settings. 
This technological shift redefined the focus of math education, emphasizing conceptual understanding over rote arithmetic skills. 
Drawing a parallel, AI-based no-code platforms present a similar transformative potential for the domain of computer science education. 
As these platforms mature, they might marginalize certain programming tasks, rendering them as obsolete. 
Consequently, it's plausible to hypothesize that computer science curricula might need to pivot, de-emphasizing certain traditional coding competencies and instead identifying a problem, defining the computational solution, and making action~\cite{Tissenbaum2019FromCT}.

Perhaps the emphasis in computational education should shift towards understanding the range of problems that programming can address. 
This would involve a deeper exploration into the art of translating real-world phenomena into quantifiable metrics, facilitating the creation of algorithmic solutions by AI-driven platforms.
Exploring these questions could lead to new approaches in computer science education, with revised assignments that are better suited to the realities of programming in the age of AI. 
This is a crucial area of research as we strive to prepare students for a future where AI is an integral part of software development.

\subsection{Can Students Efficiently Utilize AI-Based No-Code Platforms for Programming?}
The promise of these platforms making programming significantly easier might not hold entirely true.
The effectiveness of these tools depends heavily on the user's ability to precisely articulate their requirements and intentions. 
In other words, although these platforms eliminate the need for traditional coding, they introduce a new kind of complexity: the need to accurately translate one's thoughts and ideas into a detailed, unambiguous description that the AI can understand and execute.
This process of articulating precise requirements is non-trivial and involves a deep understanding of the problem at hand, a clear vision of the desired end product, and the ability to express these in a form that aligns with the AI's training and capabilities. 
This can be particularly challenging for children, who may not yet have fully developed these skills.

One emerging competency at the forefront of this evolution is educating "prompt engineering" - the art and science of crafting precise, detailed instructions for AI systems~\cite{Denny2023PromptlyUP}. 
For instance, instead of a vague request like "make me an app that translates English to Spanish," students would be trained to specify, "Create an app featuring a textbox and a button; upon entering an English sentence in the textbox and pressing the button, the system should vocalize the translated Spanish sentence." 
It also encompasses the anticipation of potential pitfalls or edge cases. 
In the aforementioned example, students would also be trained to consider scenarios like an empty textbox, leading to instructions such as, "If the textbox is devoid of content, the system should display a warning indicating the absence of input."

To fully understand the impact and potential of this shift in teaching approach, we need to examine how well younger learners can use detailed and clear language. 
We should also consider their understanding of basic programming concepts and their ability to clearly describe and communicate what they want from a program. 
By studying these aspects, we can better identify the best teaching methods to help the next generation. 
This will lead to an educational environment filled with targeted teaching strategies and tools, all aimed at helping students better understand and use these cutting-edge AI platforms.

\subsection{Can Early Exposure to No-Code Platforms Nurture Fully Capable Software Engineers?}

The advent of AI-based no-code platforms might initially seem almost magical, offering a powerful bridge between human intention and machine-executable code. 
These systems leverage statistical models trained on vast repositories of existing code, enabling them to interpolate between known data points and construct solutions that often seem remarkably optimal, especially for introductory level programming tasks.

However, beneath this seemingly magical surface lies a complex web of machine learning algorithms, which, while impressive, have limitations. As of now, these models are primarily powerful interpolators, excelling in contexts closely related to their training data. Their ability to extrapolate into new regions of the software development landscape — to generate truly novel solutions or innovate beyond their training data — remains an open question.

Moreover, as these no-code platforms become more accessible and prevalent, an important consideration emerges for computer science education. If young learners become accustomed to these AI tools, will they still develop into fully capable software engineers? Can they transition from relying on AI-generated code to creating new software solutions, understanding complex algorithms, and potentially developing innovative solutions themselves?

These questions underscore the need for careful research and thoughtful pedagogical strategies. 
As AI-based no-code platforms continue to evolve, it is crucial to investigate not only their potential but also their limitations. 
Delving deeper into the constraints of these platforms is paramount; understanding where they excel and where they falter can provide educators with a balanced view of their applicability. 
It's essential to recognize that while these tools might offer streamlined solutions for certain tasks, they might not capture the intricacies or nuances required for more complex or innovative software projects. 
By comprehensively evaluating these limitations, educators can make informed decisions on how to integrate them, ensuring that students gain both the benefits of automation and a solid grounding in foundational programming principles. 
This balanced approach aims to prepare students for a future where AI plays a pivotal role in software development, yet human insight remains irreplaceable.

\section{Conclusion}
Throughout this paper, we have underscored the foundational premise that programming, at its core, is a dialogue between human and machine. 
Historically, this conversation was predominantly one-sided, with humans making painstaking efforts to articulate their intentions in a manner comprehensible to the computer. 
This dynamic has deeply influenced the trajectory of computer science education, shaping curricula and pedagogical techniques to align with the demands of machine-centric communication.

Yet, the emergence of AI-based no-code platforms heralds a pivotal shift. 
These innovative tools are tilting the balance, enabling humans to communicate in their natural linguistic constructs and placing the onus on machines to interpret and act on these directives. 
Such a transformative change carries profound implications for the domain of computer science education.

My stance posits that, as educators, we cannot remain passive observers to this evolving landscape. 
There's an imperative to weave this changing dynamic into our curricular fabric, ensuring that future computer scientists are adept at both traditional programming and this new form of AI-mediated dialogue. 
However, this transition isn't straightforward and prompts numerous research inquiries. 
In our exploration, we've illuminated a few of these pressing questions, but this is just the tip of the iceberg. As the boundaries between human and machine communication continue to blur, our educational strategies must evolve in tandem, informed by rigorous research and a forward-looking vision.

\bibliographystyle{ACM-Reference-Format}
\bibliography{reference}

\appendix









\end{document}